\documentclass[sigconf, 10pt, nonacm]{acmart}
\usepackage{xspace}
\usepackage{makecell}
\usepackage{threeparttable}
\usepackage{listings}
\usepackage{xcolor}
\usepackage{caption}
\usepackage{subcaption}
\usepackage{placeins}
\usepackage{float}
\usepackage{enumitem}
\usepackage{needspace}
\usepackage{nopageno}  
\usepackage{times}  
\usepackage{hyperref}
\usepackage{latexsym}
\usepackage[export]{adjustbox}
\usepackage{multirow}
\usepackage{tikz}
\usepackage{array}
\usepackage{comment}
\usepackage{url}
\usepackage{xurl}
\usepackage[normalem]{ulem}
\usepackage{mdwlist}
\usepackage{algorithm}
\usepackage{booktabs}
\usepackage[noend]{algpseudocode}
\usepackage{centernot}
\usepackage{fancyvrb}
\usepackage{alltt}
\usepackage[ampersand]{easylist}
\usepackage{fancybox}
\usepackage{graphicx}
\usepackage{tcolorbox}
\usepackage{balance}
\usepackage{xcolor}
\usepackage{verbatim}
\usepackage{breakurl}
\usepackage{xcolor}
\usepackage{listings}
\usepackage{framed}
\usepackage{lipsum} % For placeholder text if needed
\usepackage[framemethod=tikz]{mdframed}

\usepackage{enumitem}
\usepackage{amsmath}

\newtcolorbox[auto counter]{policybox}[2][]{
    colback=gray!4,
    colframe=black!65,
    boxrule=0.6pt,
    arc=1.5pt,
    left=4pt,
    right=4pt,
    top=4pt,
    bottom=4pt,
    title=\textbf{Box~\thetcbcounter: #2},
    #1
}

\definecolor{codeblue}{rgb}{0,0,1}
\definecolor{codegreen}{rgb}{0,0.6,0}
\definecolor{codegray}{rgb}{0.5,0.5,0.5}
\definecolor{codepurple}{rgb}{0.58,0,0.82}
\definecolor{backcolour}{rgb}{0.95,0.95,0.92}
\definecolor{nocolor}{rgb}{1,1,1}

\definecolor{red}{rgb}{0.6,0,0} 
\definecolor{blue}{rgb}{0,0,0.6}
\definecolor{green}{rgb}{0,0.8,0}
\definecolor{cyan}{rgb}{0.0,0.6,0.6}
\definecolor{lightgray}{gray}{0.98}
\definecolor{lightblue}{rgb}{0.13, 0.67, 0.8}
\definecolor{lightorange}{RGB}{255,247,230}
\definecolor{codegreen}{rgb}{0,0.6,0}
\definecolor{codegray}{rgb}{0.5,0.5,0.5}
\definecolor{codepurple}{rgb}{0.58,0,0.82}
\definecolor{keywordcolor}{RGB}{94,20,64}
\definecolor{bluekeywords}{rgb}{0,0,1}
\definecolor{greencomments}{rgb}{0,0.5,0}
\definecolor{redstrings}{rgb}{0.64,0.08,0.08}
\definecolor{xmlcomments}{rgb}{0.5,0.5,0.5}
\definecolor{types}{rgb}{0.17,0.57,0.68}
\definecolor{KWColor}{RGB}{0,0,255}
\definecolor{AnnotationColor}{RGB}{0,137,180}
\definecolor{BlackColor}{RGB}{0,0,0}
\definecolor{CommentColor}{rgb}{0.12,0.38,0.18}
\definecolor{StringColor}{rgb}{0.06,0.10,0.98}
\definecolor{darkred}{rgb}{0.65,0,0}
\definecolor{lightgrey}{rgb}{0.8,0.8,0.8}
\definecolor{marmalade}{RGB}{193,101,18}
\definecolor{peach}{RGB}{250,217,193}
\definecolor{lime}{RGB}{220,237,193}
\definecolor{reqboxblue}{RGB}{28,35,105}
\definecolor{reqboxbg}{RGB}{248,248,255}

\newtcolorbox{requirementbox}{colback=reqboxbg,colframe=reqboxblue,boxrule=0.8pt,arc=1.2mm,left=4pt,right=4pt,top=3pt,bottom=3pt,before skip=0.6em,after skip=0.7em}

\if 0 

\definecolor{codegreen}{rgb}{0,0.6,0}
\definecolor{codegray}{rgb}{0.5,0.5,0.5}
\definecolor{codepurple}{rgb}{0.58,0,0.82}
\definecolor{backcolour}{rgb}{0.95,0.95,0.92}

\lstdefinestyle{jsonstyle}{
    backgroundcolor=\color{backcolour},
    commentstyle=\color{codegreen},
    keywordstyle=\color{blue},
    numberstyle=\tiny\color{codegray},
    stringstyle=\color{codepurple},
    basicstyle=\ttfamily\footnotesize,
    breakatwhitespace=false,
    breaklines=true,
    captionpos=b,
    keepspaces=true,
    numbers=left,
    numbersep=5pt,
    showspaces=false,
    showstringspaces=false,
    showtabs=false,
    tabsize=2,
    frame=single,
    framerule=0pt,
    rulecolor=\color{black},
    title=\lstname
}
\usepackage{framed}
\usepackage{xcolor}
\usepackage{graphicx}

\fi 

\lstdefinestyle{model-spec}{
  showspaces=false,
  showtabs=false,
  tabsize=2,
  columns=flexible,
  keepspaces=true,
  language={Java},
  numbers=left,
  xleftmargin=0pt,
  basicstyle=\ttfamily\footnotesize,
  commentstyle=\color{CommentColor}\ttfamily\footnotesize,
  stringstyle=\color{CommentColor},
  escapeinside={/*@}{@*/},
  numberstyle=\scriptsize\color{gray},
  showstringspaces=false,
  upquote=true,
  xleftmargin=1.2em,
  framexleftmargin=1.5em,
  keywords={ channels, params, signals, loop}, 
  keywords=[2]{ ms, px, kbps, s },
  keywords=[3]{ default },
  keywordstyle=\color{BlackColor}\bfseries,
  keywordstyle=[2]\color{codeblue},
  keywordstyle=[3]\color{red},
  moredelim=[il][\color{darkgray}]{$$},
}

\mdfdefinestyle{background}{backgroundcolor=lightorange,innerrightmargin=0cm,innertopmargin=-0.1cm,innerbottommargin=-0.10cm,leftmargin=+0cm, roundcorner=2pt}

\mdfdefinestyle{tracebox}{
  backgroundcolor=lightorange,
  linecolor=black,
  linewidth=0.4pt,
  roundcorner=2pt,
  innertopmargin=0.55em,
  innerbottommargin=0.55em,
  innerleftmargin=0.8em,
  innerrightmargin=1.5em,
  skipabove=0pt,
  skipbelow=0pt
}

\usepackage[framemethod=tikz]{mdframed}

\usepackage[textsize=tiny,textwidth=0.6in]{todonotes}
\newcommand{\ziqian}[1]{\todo[color=teal!20]{Ziqian: #1}}

\newcommand{\fixme}[1]{{\bf\textcolor{red}{[#1]}}}

\newcommand{\sys}{\textsc{Conflux}\xspace}

\makeatletter
\renewcommand\@formatdoi[1]{\ignorespaces}
\makeatother

\makeatletter
\def\@copyrightspace{\relax}
\makeatother

\AtBeginDocument{%
  }

\setcopyright{none}
\acmISBN{}
\acmYear{}
\acmDOI{}

\newcommand{\para}[1]{\smallskip\noindent {\bf #1} }
\newcounter{defn}[section]
\renewcommand{\thedefn}{\arabic{section}.\arabic{defn}}

\title{The Model in the Middle: Toward AI-Native Real-Time Communication}

\begin{abstract}
\if 0
omni-model serving is a combine of three dimensions: network; serving framework and model.
this is a dedicate framework because it changes: 1. more real-time requirement; 2. workload changes to streaming, so more component outside the model: flow-matching and chunk; 
we need a gym to testing/generating trace to this new model: interactively simulating user behavior to react the model output.

new insight: we need a new vllm-omni-realtime, which connect vllm inference and websocket/webrtc transport, it has abstraction of chunk transport/schedule.
\fi

Full-duplex omni models are transforming human--AI interaction from turn-based exchanges into continuous multimodal conversations in which speaking, listening, and reasoning unfold concurrently. Rather than viewing the model as a replacement for a human endpoint, we argue for a new perspective: the model is a stateful computational middlebox inside a human-centered feedback loop, with network transport, model serving, and user playback jointly shaping how the interaction evolves. This perspective breaks the traditional boundaries among stages designed around local objectives. Rather than optimizing them in isolation, an AI-native real-time stack should allow the state of each stage to shape the actions of the others. We explore three cross-stage coordination opportunities: network-aware inference scheduling, execution-aware transport prioritization, and playback control that accounts for both network and model variability. We are building \sys to explore these ideas, and preliminary results show substantial improvements in response latency and playback deadline adherence under network degradation. More broadly, we call for an AI-native real-time communication stack that resolve the joint control problem spanning communication, computation, and playback.

\if 0
Full-duplex omni models are transforming human--AI interaction from turn-based exchanges into continuous multimodal conversations in which speaking, listening, and reasoning unfold concurrently.
Rather than viewing the model as a replacement for a human endpoint, we argue for a new perspective: the model is a stateful computational middlebox inside a human-centered feedback loop, with network transport, model serving, and user playback jointly shaping how the interaction evolves. This perspective breaks the traditional boundaries among stages designed around local objectives. Rather than optimizing them in isolation, an AI-native real-time stack should allow the state of each stage to shape the actions of the others.
In this paper, 
we leverage this perspective to explore three
cross-stage coordination opportunities:
network-aware inference scheduling, model-aware transport prioritization, and playback
control that accounts for both network and model variability. We are building \sys to explore these ideas, and our
preliminary results show substantial gains in user quality of experience. 
More broadly, we call for an
AI-native real-time communication stack that jointly controls communication,
computation, and playback around the evolving interaction.
\fi
\end{abstract}

\author{Ziqian Liu}
\affiliation{
  \institution{The University of Hong Kong}
  \city{}
  \country{}
}

\author{Minghao Li}
\affiliation{
  \institution{The University of Hong Kong}
  \city{}
  \country{}
}

\author{Yiming Qiu}
\affiliation{
  \institution{The University of Hong Kong}
  \city{}
  \country{}
}

\begin{document}

\maketitle

\section{Introduction}

\begin{figure}[t]
  \centering
  \includegraphics[width=0.46\textwidth]{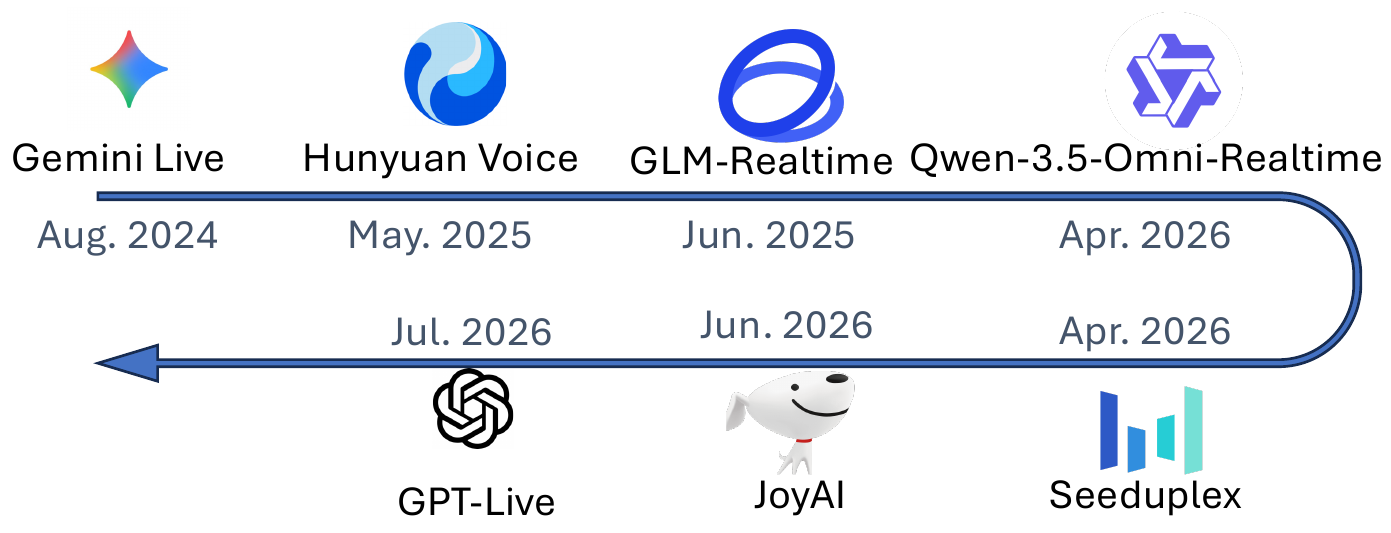}
  \vspace{-4mm}
  \caption{Real-time omni-model development.}
  \vspace{-5mm}
  \label{fig:intro1}
\end{figure}

\begin{figure}[t]
  \centering
  \includegraphics[width=0.48\textwidth]{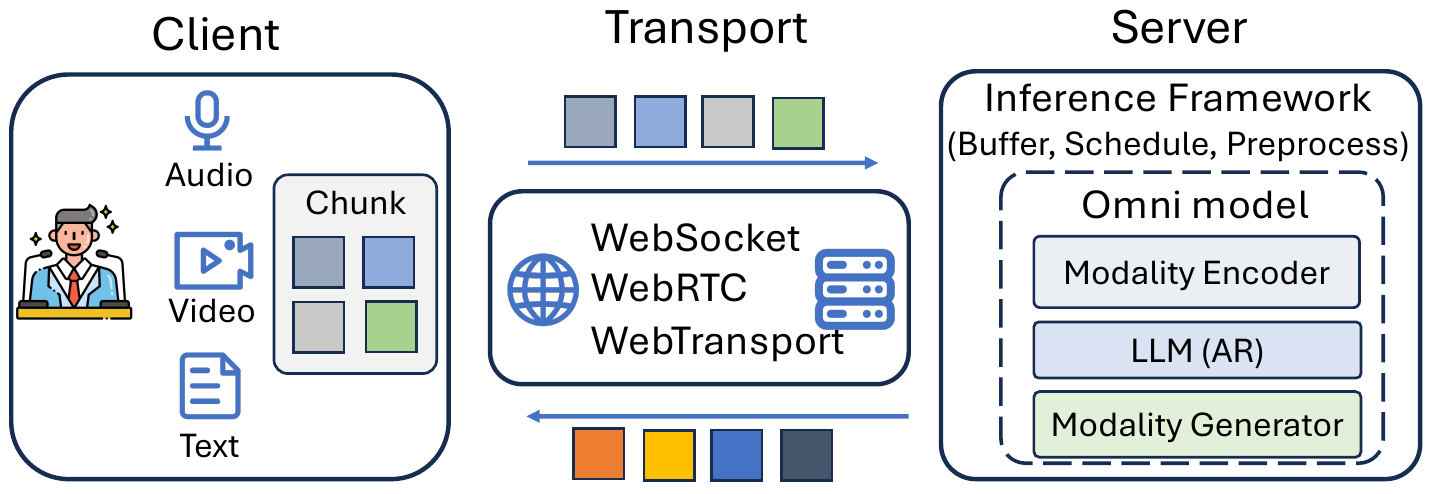}
  \vspace{-6mm}
  \caption{Full-duplex omni-model workflow.}
  \vspace{-5mm}
  \label{fig:intro2}
\end{figure}

Large language models are rapidly moving beyond turn-based question answering toward real-time, full-duplex multimodal interaction~\cite{doubao2026,microsoft_copilot_vision_2025,soulfun_ai_2026,duolingo_max_2026,glm_realtime,grok2026}. Instead of submitting a complete query and waiting for a response, users can speak while sharing their camera or screen, interrupt the model mid-sentence, point to new visual evidence, and continuously refine their intent as the interaction unfolds. A growing landscape of systems, as shown in Figure~\ref{fig:intro1}, including Gemini Live~\cite{google_gemini_live}, Qwen-3.5-Omni-Realtime~\cite{team2026qwen3}, and GPT-Live~\cite{openai2026gptlive}, is beginning to support this experience. Their capabilities enable applications ranging from visual assistance~\cite{doubao2026,chatgpt2026,inmo2026} and real-time translation~\cite{translation1,translation2,translation3} to meeting agents~\cite{meeting1,meeting2}, interactive avatars~\cite{avatar1,avatar2,avatar3}, and embodied AI~\cite{embodied1,embodied2}. More fundamentally, they turn model interaction into a continuous real-time process in which perception, reasoning, generation, and user feedback overlap in time~\cite{yao2026joyai,openai2026gptlive}.

Conventional real-time communication~\cite{webrtc,websocket,webtransport} and LLM serving~\cite{vllm,sglang,vllm-omni} both operate under a sender–receiver abstraction: one endpoint produces information and the other consumes it, whether that receiver is human or AI. Full-duplex omni-model interaction calls for a fundamentally different perspective. As Figure~\ref{fig:intro2} shows, a full-duplex model continuously updates its state from streaming chunks while producing corresponding output chunks from that evolving state, forming a latency-sensitive uplink–model–downlink path. As such, we argue that a full-duplex model is best understood as a \emph{stateful computational middlebox inside a human-centered feedback loop}. The central shift is not that one receiver becomes AI, but that communication, computation, and user reaction now co-evolve as a single closed-loop system.

This closed-loop structure breaks the traditional boundaries among network transport, model serving, and user playback. These components were designed around different local objectives: transport delivers media efficiently, model serving manages inference latency and throughput, and playback absorbs jitter to maintain smooth output. In full-duplex interaction, however, each component changes the operating conditions faced by the others. Network delay determines when input becomes available to the model; model serving determines when chunks enter the downlink; and the model’s modality, fidelity, and freshness requirements constrain what the network must deliver. Playback must then absorb variability introduced by both the network and the model. 
Full-duplex AI thus shifts the design focus from optimizing individual stages to jointly optimizing their interactions across the end-to-end path.

In this paper, we envision an AI-native real-time communication stack that accounts for the interactions among network transport, model serving, and user playback. Unlike recent systems that optimize communication via costly model context and task semantics~\cite{chen2026bits,wu2025chat}, we argue that broad cross-stage coordination can be achieved through execution-level properties. 
%\ziqian{more text is needed for why context-agnostic}
Our central idea is that chunks provide a common coordination substrate across the end-to-end path. As a chunk progresses through transport, inference, and playback, it can carry and accumulate cross-stage signals such as the remaining latency budget, while preserving a stable identity. 
We therefore call for chunk-level interfaces that expose the system conditions and requirements needed for coordination. Such interfaces allow stages to jointly determine when each chunk should be transmitted, processed, and presented, without requiring them to understand what the chunk means.

Our first observation is that network conditions can change the urgency and usefulness of model-serving work. Conventional serving schedulers make admission and batching decisions largely from server-local signals, such as queuing time, request age, and GPU efficiency. In full-duplex interaction, however, chunks reach the model after accumulating different amounts of network delay, leaving them with different remaining latency budgets. A delayed barge-in or turn-transition chunk may require immediate processing before the model produces more obsolete output, whereas a visual update may already be too stale to justify further queuing or computation. 
%The chunk-level interface should allows the runtime to account for this accumulated delay without interpreting the chunk’s content. This motivates network-aware model serving that admits latency-critical chunks immediately, batches only when waiting is safe, and degrades or discards work whose useful window has passed. 
Model serving should therefore schedule chunks by their remaining value to the interaction, not merely by their arrival order or batching efficiency.

Conversely, our second observation is that the model can expose how input chunks should compete for network resources. We argue that transport prioritization need not depend on online semantic interpretation~\cite{chen2026bits,wu2025chat}, because full-duplex chunks already differ in execution-level properties: some are urgent or irreplaceable, some require reliable delivery, while others can be degraded, deferred, or replaced by fresher inputs. The model runtime can expose these properties as a transport specification, which the sender combines with dynamic signals such as deadline slack and network conditions to prioritize chunks across modalities and channels. This allows barge-in, speech, control traffic, and visual context to take precedence over one another without requiring the transport to infer what any chunk means. The opportunity is therefore to make network transport execution-aware without making it semantics-aware.

Our third observation is that full-duplex AI presents a new playback decision: when should each model response begin? Conventional real-time communication maintains an ongoing media timeline, with buffering primarily used to absorb network jitter.  A full-duplex model, in contrast, alternates between listening and producing finite responses. Each time a response starts, the receiver must establish a new playback timeline from the first output chunk. Starting immediately minimizes time to first response, but leaves little protection if subsequent chunks are delayed; waiting creates playback slack, but directly slows the interaction. Moreover, future arrivals depend not only on downlink conditions, but also on model serving progress. AI-native playback must therefore decide when each response is ready to begin, absorbing variability across the entire uplink-model-downlink path.

We are building \sys, an AI-native real-time communication system that instantiates these observations across transport, model serving, and playback stages. We describe its design principles and provide initial evidence of its potential.
%We are developing \sys, an AI-native real-time communication system that operationalizes these insights across the network transport, model serving, and user playback layers. We describe its design principles and provide initial evidence of its potential.
\if 0
Our second observation is that conversely, model behavior should shape how input chunks are transported. Conventional media transport adapts fidelity and reliability primarily for human perception, whereas full-duplex uplink streams are consumed by a model with modality-specific operating requirements. 
More data is not always better: once an input reaches sufficient quality for model consumption, additional fidelity may increase transmission and inference pressure without improving the response~\cite{wu2026artic}. We therefore propose that models expose a transport specification describing the timing, fidelity, reliability, and alignment ranges within which their inputs remain useful. The network can then adapt chunk size, modality quality, delivery channel, or path within this declared operating envelope as conditions change. Importantly, this interface need not reveal which object, region, or utterance is semantically important: transport only needs to satisfy the model’s stated consumption requirements. \emph{Transport should therefore be defined by how the model consumes information, without being required to understand what that information means.}
\fi

\section{Motivation}
% \fixme{rewrite for more vision}
% 2.1: delete details and talk about scenario: what we use full-duplex model for ai assistant, merge now 2.1 and 2.2 for background and give one sentence to sum up, or move figure 2 to intro

% 2.2: Case study bargein, network-conditioned model behavior to show network is important

% 2.3: model-centric requirement for multi-modality and opportunity to leverage multi channel and multi path

% 2.4: problem formulation and receiver side opportunity
% \begin{figure}[t]
%   \centering
%   \includegraphics[width=0.5\textwidth]{Figures/2.1.png}
%   \caption{Full-duplex Omni-model workflow.}
%   \label{fig:2.1}
% \end{figure}

\subsection{Transport Pattern in Full-Duplex AI}

\begin{figure}[t]
  \centering
  \includegraphics[width=0.5\textwidth]{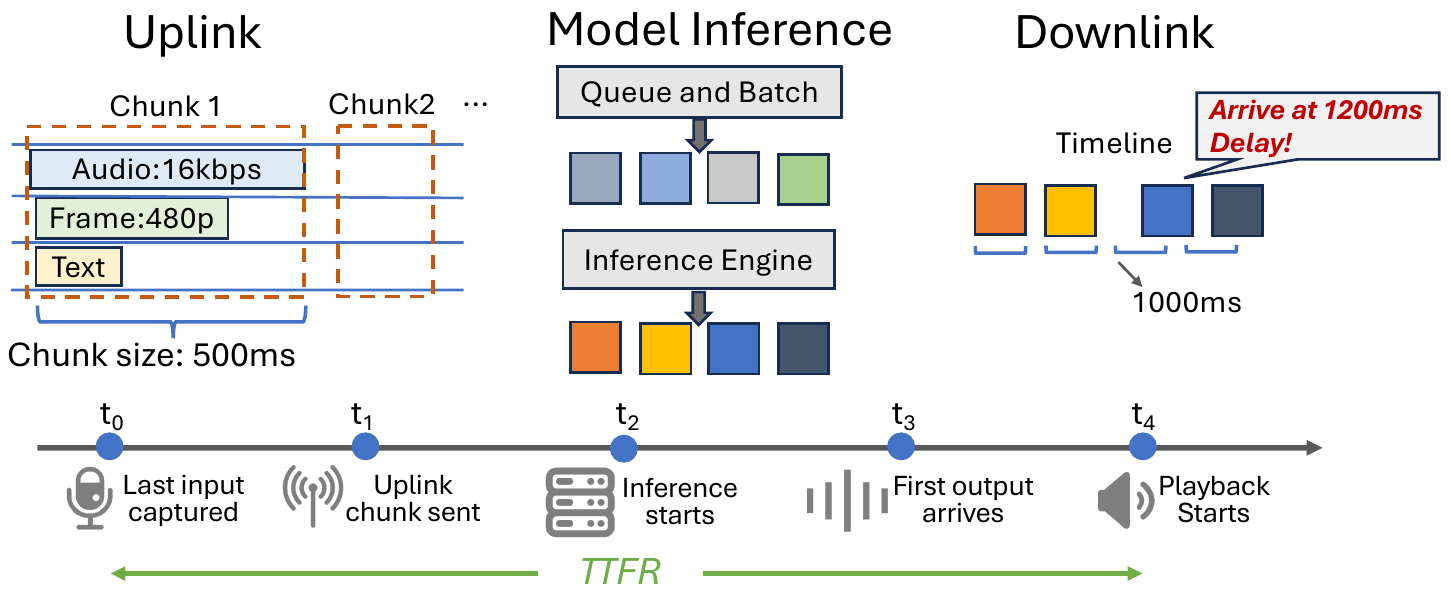}
  \vspace{-6mm}
  \caption{Full-duplex omni-model transport flow.}
  \vspace{-4mm}
  \label{fig:2.1}
\end{figure}

Full-duplex omni-models treat chunks as the basic transport objects. 
As shown in Figure~\ref{fig:2.1}, each uplink chunk represents a bounded interaction interval, e.g., a 500~ms window, and contains multiple aligned modalities, such as speech samples at a given bitrate, selected video frames with configurable resolution and frame rate, text, and control events.
These chunks are transmitted over the uplink using transport channels such as WebRTC or WebSocket.
Then, each input chunk proceeds through the server-side model pipeline, where it is preprocessed, encoded, and batched.
The autoregressive LLM subsequently samples an interaction action: \textit{silence}, producing no media output; \textit{delegate}, invoking a background task; or \textit{response}. A \textit{response} action generates a token sequence, which is rendered by modality-specific generators, such as TTS, and packaged into a sequence of output chunks for downlink delivery. This process establishes a logical correspondence between each input chunk and the resulting output chunks.
The receiver maintains a playback timeline anchored by the first output chunk, with subsequent deadlines determined by chunk duration and playback buffer. Chunks arriving before their deadlines enable continuous playback, whereas late arrivals introduce gaps and user-perceived glitches.

\subsection{Network-Conditioned Model Behavior}

\begin{figure}[t]
  \centering
  \includegraphics[width=0.47\textwidth]{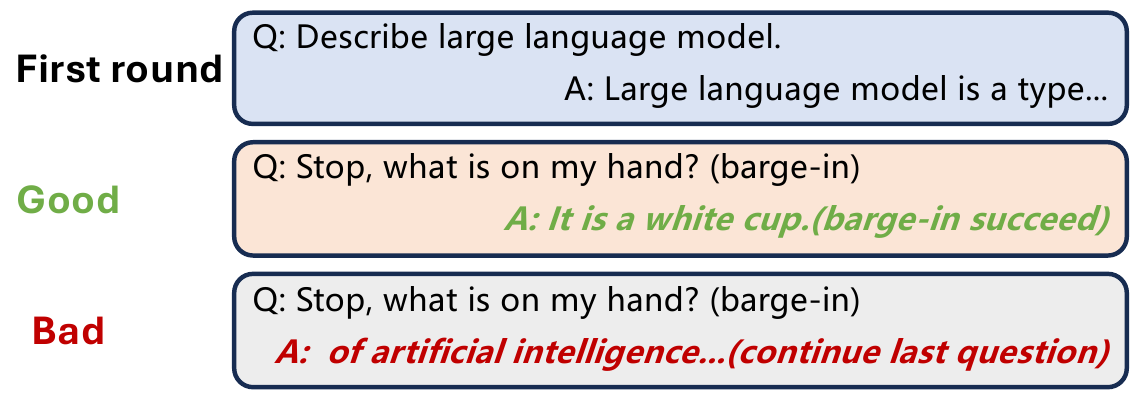}
  \vspace{-4mm}
  \caption{A barge-in example.}
  \vspace{-4mm}
  \label{fig:bargein}
\end{figure}

\begin{table}[t]
\centering
\caption{Barge-in latency by network condition.}
\vspace{-4mm}
\label{tab:bargein-latency}
\small
\setlength{\tabcolsep}{4pt}
\begin{tabular}{lcccc}
\toprule
\textbf{Metric} & \textbf{Good} & \textbf{Constrained} &
\textbf{Limited} & \textbf{Bad} \\
\midrule
Bandwidth (Mbps) & 20 & 10 & 5 & 2.5 \\
RTT (ms) & 20 & 40 & 80 & 160 \\
Jitter (ms) & 0 & 5 & 10 & 20 \\
Barge-in latency (ms) & 542.7 & 686.6 & 1202.9 & 5040.9 \\
\bottomrule
\end{tabular}
\vspace{-4mm}
\end{table}

\if 0
In traditional media systems, poor network conditions mainly appear as receiver-side quality degradation, such as stalls, artifacts, or lower resolution. In full-duplex AI media, however, the same network conditions can change the model's input context before generation happens. A delayed, degraded, or misaligned chunk may still be delivered, but it may no longer represent the interaction state that the model is currently reasoning about. This makes the effect of the network semantic rather than purely perceptual: delay determines whether an input is still timely, loss and congestion determine whether the model receives enough evidence, and buffering or reordering determines whether different modalities are interpreted together correctly. As a result, the transport layer does not merely affect how smoothly the user receives media; it affects what the model knows, when the model knows it, and whether the model's next output is grounded in the right context. Therefore, full-duplex AI media should evaluate transport decisions by their impact on model utility, not only by packet delivery or playback quality. 
\fi

%\minghao{We need more background about how "barge-in" is implemented? Barge-in is a signal that needs high priority.}
\noindent In traditional media systems, a poor network condition mainly appears as receiver-side quality degradation, such as stalls and lower resolution~\cite{meng2024latency,meng2026mae,shen2026mortise,rtc4,bwe-challenge-mmsys24}. In full-duplex AI media, however, it can affect what content is produced in the first place, because delayed input changes the model state from which subsequent outputs are generated. 
%A barge-in is a state-changing control signal that invalidates the current response. 
%Rather than simply stopping transmission, the 
Take user barge-in as an example, the signal must enter the model’s inference loop so that the model can update its interaction state and decide to stop speaking and listen to the new input.  
%As barge-in determines the validity of model output, it should be delivered and processed with high priority.
As shown in Figure~\ref{fig:bargein}, after the model starts answering the first question, the user interrupts with a new visual question. Under a good network condition, the model receives the interruption in time and switches to the new responses, while under network degradation, the delayed barge-in causes it to continue the obsolete response. Table~\ref{tab:bargein-latency} shows this case quantitatively in MiniCPM-o-4.5 Demo~\cite{cui2026minicpm} with voice activity detection (VAD~\cite{vad}) setup: as network condition degrades, barge-in latency increases from 542.7~ms to 5040.9~ms.

To ensure the timely processing of interaction-sensitive inputs, full-duplex model serving needs to become inherently network-aware.
%Once a state-changing signal reaches the server, the runtime should prioritize the affected session to terminate obsolete generation and reduce interaction latency. 
The inference scheduler should maintain lightweight per-session information, 
such as speaking state, input queuing delay, and, crucially, network conditions. 
Using these signals, the runtime can determine whether a chunk should be admitted immediately, preempt ongoing work, or tolerate a short delay to improve batching efficiency. 
In particular, network and session state reveal whether additional waiting would merely increase latency or further reduce the usefulness of ongoing generation. Therefore, full-duplex AI media should schedule model inference in a network- and session-aware manner rather than relying solely on request arrival order or GPU utilization~\cite{serving1,serving2,serving3}.

\begin{requirementbox}
\textbf{Insight 1:} Model inference should be per-session and network-aware to optimize end-to-end interaction quality.
\end{requirementbox}

\subsection{Model-Defined Transport Requirements}
\if 0
\begin{table}[t]
\centering
\caption{Model-defined transport requirements}
\label{tab:model-req}
\small
\setlength{\tabcolsep}{3pt}
\begin{tabular}{p{0.13\linewidth} p{0.86\linewidth}}
\toprule
\textbf{Modality} & \textbf{Requirement} \\
\midrule
Control & Timestamped event; strict freshness; no invalid-state loss \\
\midrule
Audio & Speech chunks; bitrate; delay-sensitive; lossy tolerance \\
\midrule
Video & Selected frames; resolution; ROI priority \\
\midrule
Text & Tokens; exact fidelity; reliable delivery for critical content \\
\bottomrule
\end{tabular}
\end{table}
\fi

\begin{table}[t]
\centering
\caption{Modality-specific transport characteristics.}
\vspace{-4mm}
\label{tab:model-req}
\small
\setlength{\tabcolsep}{4pt}
\begin{tabular}{lcccc}
\toprule
\textbf{Modality} &
\textbf{Urgent?} &
\textbf{Reliable?} &
\textbf{Replaceable?} &
\textbf{Adaptive?} \\
\midrule
Control & High & High & No  & No \\
Audio   & High & Medium & No  & Yes \\
Video   & Medium & Low & Yes & Yes \\
Text    & Medium & High & No  & No \\
\bottomrule
\end{tabular}
\vspace{-4mm}
\end{table}

Recent work~\cite{wu2025chat,wu2026artic,meng2026sema} shows that model accuracy does not improve monotonically with input volume: insufficient fidelity can cause inference errors, while additional data provides little benefit beyond a sufficient quality level. However, identifying this boundary from the semantic context of each request requires task-specific online analysis and is costly and difficult to operationalize in transport; for example, CLIP-based ROI identification incurs 27.13\% additional inference cost~\cite{wu2026artic,clip}.
Our key observation is that multimodal inputs already have distinct and well-defined roles in model interaction. As shown in Table~\ref{tab:model-req}, control events, audio, video, and text differ in urgency, reliability, replaceability, and adaptability. For example, control events and speech directly affect turn-taking, visual frames provide replaceable grounding evidence, and text often carries exact symbols that require reliable delivery.

Omni-models should therefore expose explicit modality-specific transport requirements, including minimum fidelity, tunable operating ranges, latency sensitivity, reliability, and cross-modal alignment constraints. The transport layer can then satisfy the minimum requirements needed for accurate inference while avoiding unnecessary transmission, and independently map each modality to suitable channels or network paths rather than applying a uniform policy to all inputs.
The resulting objective is an accuracy-bounded performance optimization: deliver sufficient model-useful evidence within its validity window while minimizing transmission overhead and end-to-end delay.

\begin{requirementbox}
\textbf{Insight 2:} Omni-models should expose modality-specific requirements to enable model-defined transport.
%multiple channels and multiple paths selection.
\end{requirementbox}

\subsection{Interaction QoE and Playback Deadlines}
\label{sec:qoe}

\noindent Traditional real-time video transport optimizes continuous media streams, where receiver-side buffering primarily absorbs packet-arrival variation and maintains smooth playback~\cite{carlucci2016analysis,rtc1,rtc2,rtc3,mowgli-nsdi25}. AI-native real-time communication exposes a different optimization space: model outputs arrive as discrete response chunks whose timing depends on both network conditions and model execution. User-perceived quality therefore depends on how quickly a response begins and whether later chunks arrive before their playback deadlines.
We characterize interaction QoE using two metrics: \emph{Time to First Response} (TTFR), which measures the delay until the first useful output reaches the user, as shown in Figure~\ref{fig:2.1}, and \emph{Deadline Miss} (DM), which captures the accumulated delay time of chunks that miss their playback deadlines. 
%\minghao{maybe this could be min over t and TTFR, DeadlineMiss can be a function over time or workload.}

\if 0
This start time anchors the response timeline, and subsequent chunk deadlines follow from their media duration and stream order. A smaller buffer reduces TTFR, whereas a larger buffer gives later chunks more slack and reduces DM.
Unlike the jitter buffer of a continuous stream, this waiting time can be chosen independently for each model response. By considering network conditions, model backpressure, and recent interaction latency, the receiver can start earlier under stable conditions and reserve more buffering when later chunks are likely to be delayed.
\fi

After the first chunk arrives, the receiver waits for a buffer interval before starting playback.
Let $t_0$ denote the arrival time of the first output chunk~$0$, $b_0$ the initial playback buffer, and $d_i$ the media duration of chunk~$i$. Playback begins at time $t_0+b_0$, which anchors the response timeline. Accordingly, the playback deadline of chunk~$1$ is $t_0 + b_0 + d_0$.
A smaller $b_0$ reduces TTFR, whereas a larger $b_0$ provides more slack for subsequent arrivals and reduces DM. Unlike the jitter buffer of a continuous stream, $b_0$ can be selected independently for each model response based on network conditions, model backpressure, and recent interaction latency.

\begin{requirementbox}
\textbf{Insight 3: }Playback should be scheduled based on network delay and model backpressure to maximize QoE.
\end{requirementbox}

\section{Research Agenda}

\begin{figure}[t]
  \centering
  \includegraphics[width=0.48\textwidth]{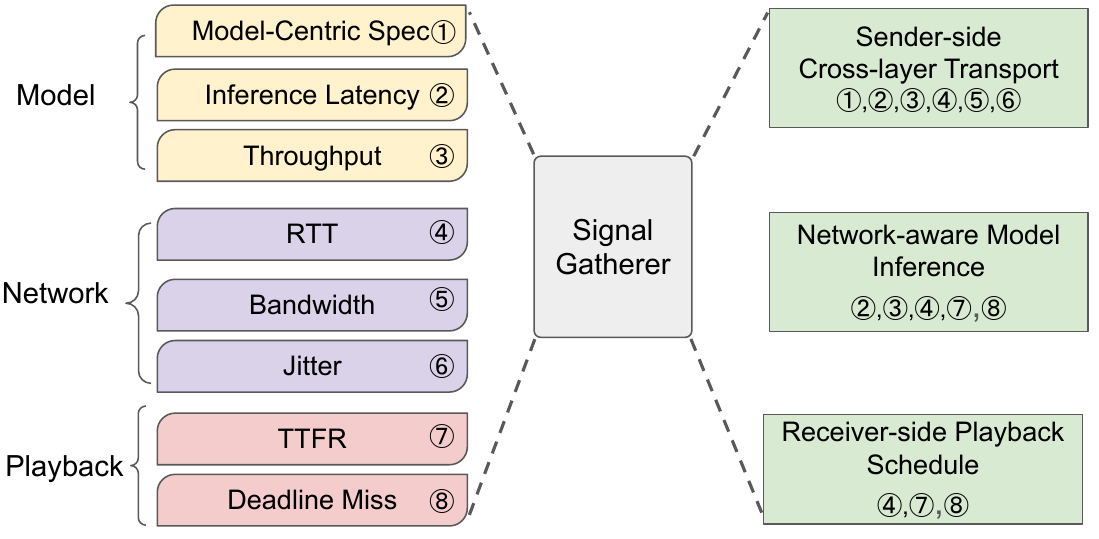}
  \vspace{-7mm}
  \caption{\sys Overview.}
  \vspace{-5mm}
  \label{fig:overview}
\end{figure}

As illustrated in Figure~\ref{fig:overview}, \sys uses model, network, and playback signals to coordinate three runtime actions. Model signals expose modality requirements and inference backpressure; network signals such as RTT, bandwidth, and jitter estimate chunk delivery conditions; and playback signals such as TTFR and DM reflect interaction quality. Guided by these signals, the sender adapts chunk size, modality channels, fidelity, and reliability; the server prioritizes chunks for immediate inference, batching, or degradation based on urgency and staleness; and the receiver adjusts the first-chunk playback buffer using end-to-end latency. Together, these coordinated actions form an end-to-end control loop that adapts transport, inference, and playback to the evolving interaction state.

\begin{figure}[t]
    \centering
    \begin{mdframed}[style=background]
\begin{lstlisting}[basicstyle=\footnotesize\ttfamily, numbers=left]
channels:
  audio:
    protocol: WebSocket
    direction: uplink downlink
params:
  audio_bitrate:
    channel: audio
    range: 8 48 default 24 kbps
  video_resolution:
    channel: video
    values: 240 360 480 720 default 480 px
  chunk_size:
    values: 0.5 1 2 default 1 s
signals:
  server_tbt_threshold: 40 ms
  server_ttft_threshold: 120 ms
loop:
  policy: delay-gradient-aimd
  interval: 1000 ms
\end{lstlisting}
    \end{mdframed}
    \vspace{-4mm}
    \caption{Example model-centric transport specification.}
    \vspace{-6mm}
    \label{fig:model-centric-transport-spec}
\end{figure}

\subsection{Sender-Side Cross-Layer Transport}
\label{sec:sender-side-transport}

The sender is the first control point in the full-duplex AI media loop. It observes cross-layer signals before deciding how to schedule outgoing chunks and coordinate multiple modality channels. 
These signals include the model specification, which defines the input requirements, as well as runtime signals from network and inference engine that indicate whether the current input path can deliver and process chunks timely.

As shown in Figure~\ref{fig:model-centric-transport-spec}, the specification exposes the transport configuration available to the sender, including media channels and protocols, tunable parameters, runtime feedback signals, and the adaptation policy.
The \textit{channels} field declares the logical media channels, their transport protocol, and direction. The \textit{params} field defines configurable knobs, such as audio bitrate, video resolution, and chunk duration, together with their valid ranges or candidate values and default settings. 
It can also define an adaptation order, allowing the sender to reduce less critical dimensions first, such as lowering video resolution before audio bitrate when speech freshness and intelligibility are more important than visual fidelity.
The \textit{signals} field combines network measurements, including RTT, jitter, loss, and available bandwidth, with server-side model backpressure indicators such as chunk serving latency, processing throughput, and TTFT.
%The \textit{signals} field combines client-side feedback, network measurements, and model backpressure. Client-side TTFR reflects interaction responsiveness; RTT, jitter, loss and bandwidth characterize transport conditions; and chunk serving latency, processing throughput, and TTFT indicate back pressure of inference engine. 

At runtime, the sender applies the configured policy to translate the specification and cross-layer observations into a joint transmission decision across modality channels. The decision determines which chunks should be degraded, deferred, or discarded based on their latency sensitivity, deadline slack, and relevance to the current interaction, allowing speech, barge-in, and control events to take precedence over less urgent visual content. The policy can range from lightweight threshold-based or delay-gradient AIMD controllers to model predictive control or learned policies, depending on the available system model and runtime data.

\if 0
The inference engine further uses chunk metadata to prioritize serving requests across concurrent sessions. Each incoming chunk carries session-level timing metadata, including the estimated end-to-end latency accumulated across the uplink, model stage, and downlink. The model-side scheduler uses this latency as a priority signal: chunks from sessions with higher end-to-end latency are promoted, because delaying them further is more likely to miss the interaction deadline or increase response lag. In contrast, chunks from sessions with lower accumulated latency can be served later, allowing inference capacity to focus on sessions whose delay is more likely to affect full-loop QoE.
\fi

\subsection{Network-Aware Model Inference}

\noindent 
Full-duplex AI media requires network-aware inference scheduling that accounts for each session's interaction state and end-to-end latency.
The key question is not only whether delaying a chunk improves serving efficiency, but whether it changes the chunk's meaning or usefulness. The scheduler therefore maintains lightweight per-session context, including duplex phase, turn and interaction states, recent QoE, network quality, output progress, and inference backlog. Each chunk is  scheduled based on these states before admission, allowing the runtime to prioritize state-changing events while batching less urgent inputs for better throughput.

\begin{policybox}[label={box:state-aware-scheduling}]
{State-Aware Inference Scheduling Policy.}
\begin{enumerate}[leftmargin=1.2em, label=(\arabic*), itemsep=2pt, topsep=2pt]
    \item 
    Update network, QoE, and inference states.

    \item 
    For each pending chunk $c$ in session $s$,
    \[
    P(s,c)=PriorityFunc(I(s,c)+Q(s)+N(s)),
    \]
    $P$: Priority, $I$: Interaction state, $Q$: QoE, $N$: Network.

    \item Sum over priority of current batched chunks $C$ and select action $A$ based on threshold $\theta$.
    \[
    A(s)=
    \begin{cases}
    \textsc{AdmitNow}, & \sum_{c \in C}P(s,c)\ge\theta,\\
    \textsc{WaitBatch}, & \sum_{c \in C}P(s,c)<\theta,\\
    \end{cases}
    \]

    \item  Commit the batch when it is full, its wait limit expires, or any queued chunk becomes urgent.

\end{enumerate}
\end{policybox}

At each scheduling epoch, the scheduler refreshes active session states and assigns each pending chunk a priority. Box~\ref{box:state-aware-scheduling} summarizes how the scheduler converts this per-session state into priority-based inference actions. The priority score captures whether the chunk may change the model state, such as a turn transition, barge-in event, or control signal. QoE ensures sessions that have recently experienced poor interaction quality, such as high end-to-end latency, delayed responses, stalls, or failed interruptions, will be given high priority, while network degradation captures unstable transport conditions that make additional waiting more harmful. The scheduler aggregates the priorities of queued chunks within each candidate batch and uses the resulting batch priority to determine when inference should begin. If the aggregate priority exceeds threshold, the batch is admitted immediately, allowing urgent interaction events and sessions with poor QoE or degraded network conditions to shorten the batching delay. Otherwise, the batch waits to improve serving efficiency.

\subsection{Receiver-Side Playback Scheduling}
\label{sec:design:receiver-playback}

The receiver controls the last stage of the full-duplex loop by setting the playback buffer duration after the first output chunk arrives and before playback begins.
This decision balances responsiveness against playback smoothness: a smaller buffer reduces time to first response, whereas a larger buffer gives subsequent chunks more time to arrive before their playback deadlines. 
Network conditions and model backpressure determine end-to-end latency, shaping subsequent chunk arrivals relative to their playback deadlines.
Therefore, at the beginning of each response, the reciever derives a congestion signal from recent playback and delivery dynamics. Based on these observed signals, the receiver adapts the playback buffer using a configurable congestion-control-inspired policy.
The receiver then updates the buffer time $b_t$ at time $t$ as:
\[
b_t =
\begin{cases}
\min\left\{\gamma b_{t-1},\, b_{\max}\right\}
& L_t > \phi_{\text{high}}, \\[3pt]
\max\left\{b_{t-1}-\delta,\, b_{\min}\right\}
& L_t < \phi_{\text{low}}, \\[3pt]
b_{t-1}
& \text{otherwise},
\end{cases}
\]
where $L_t$ denotes the observed end-to-end latency, $\gamma>1$ is the multiplicative increase factor, $\delta>0$ is the additive decrease step, and $\phi_{\text{high}}>\phi_{\text{low}}$ define a hysteresis region in which the buffer remains unchanged. When $L_t$ exceeds $\phi_{\text{high}}$, the receiver rapidly increases the buffer to absorb larger inter-chunk arrival gaps. When $L_t$ falls below $\phi_{\text{low}}$, it gradually decreases the buffer to improve responsiveness. Values between $\phi_{\text{low}}$ and $\phi_{\text{high}}$ do not trigger an adjustment, preventing small latency fluctuations from causing oscillation. For different models and deployment environments, the input signal can incorporate additional metrics, such as network jitter, to enable faster and more sensitive adaptation to short-term delivery changes.

\section{Preliminary Validation}
\begin{table}[t]
\centering
\caption{Experiment network settings.}
\label{tab:network-profiles}
\small
\setlength{\tabcolsep}{8pt}
\begin{tabular}{lcccc}
\toprule
\textbf{Metric} &
\textbf{L1} &
\textbf{L2} &
\textbf{L3} &
\textbf{L4} \\
\midrule
Bandwidth (Mbps) & 10.0 & 5.0 & 2.5 & 1.5\\
RTT (ms)       & 40  & 80  & 160  & 200 \\
Jitter (ms)      & 0   & 10  & 20  & 40 \\
\bottomrule
\end{tabular}
\vspace{-6pt}
\end{table}

\para{Experiment settings.}
We conducted a preliminary experiment using MiniCPM-o-4.5~\cite{cui2026minicpm}, an open-source full-duplex omni-modal model, deployed on an AWS instance with one NVIDIA L40S GPU (48~GB VRAM). The client runs on another cloud provider, while both the client and the server are located in the US East region and communicate over the public Internet. We use the MiniCPM demo repository as baseline and emulate the four network profiles in Table~\ref{tab:network-profiles} using Linux Netem, ranging from high-bandwidth, low-latency conditions to severely constrained links. For all profiles with nonzero jitter, we set the jitter correlation to 25\%.

\begin{figure}[t]
  \centering
  \begin{subfigure}{0.46\textwidth}
    \centering
    \includegraphics[width=\linewidth]{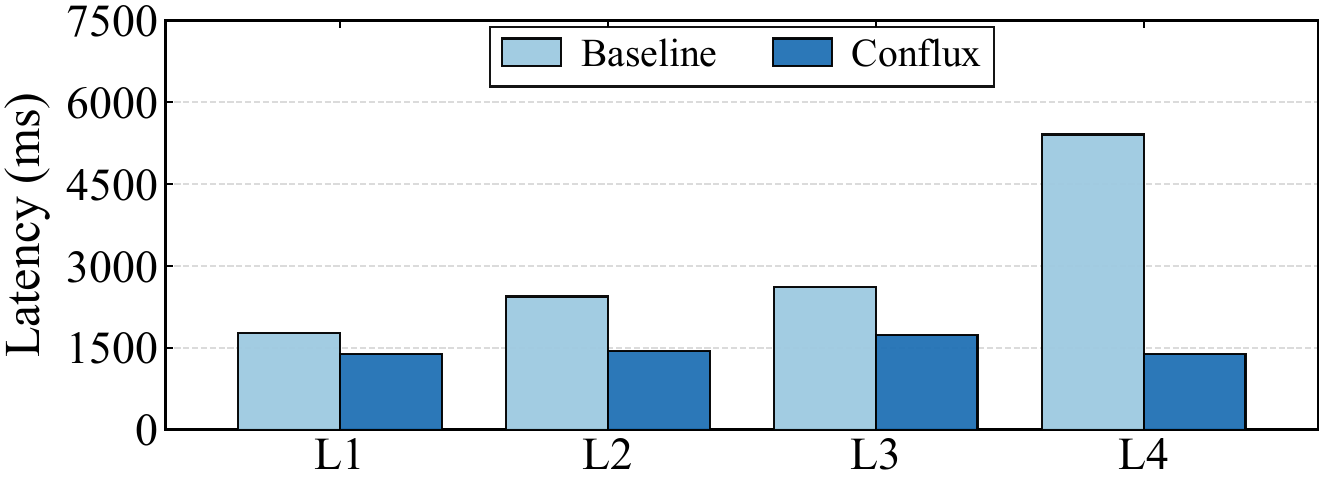}
    \vspace{-4mm}
    \caption{Time to first response.}
    \label{fig:ttfr}
  \end{subfigure}

  \begin{subfigure}{0.46\textwidth}
    \centering
    \includegraphics[width=\linewidth]{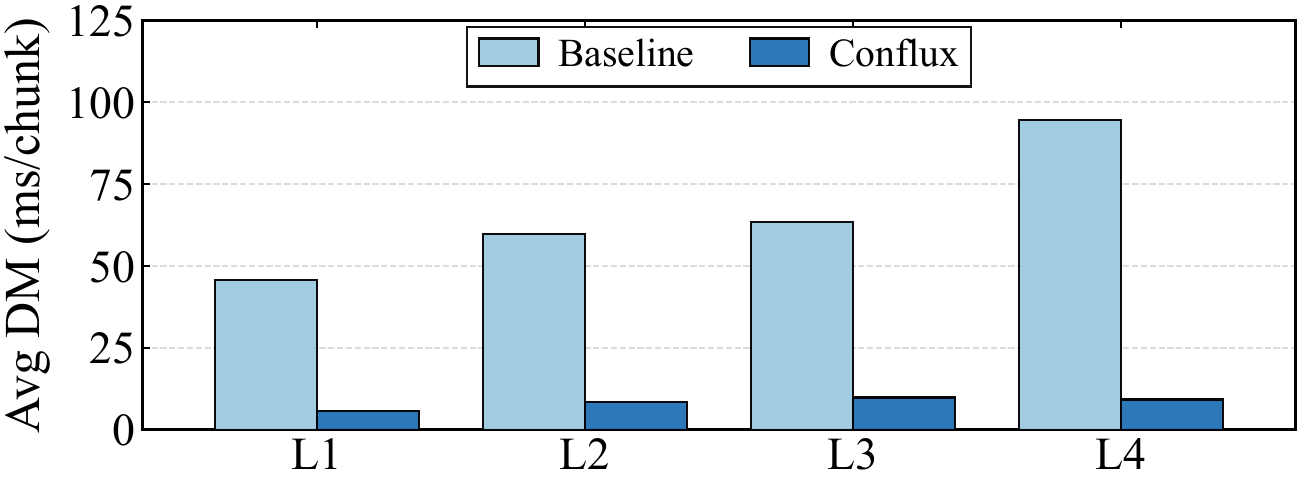}
    \vspace{-4mm}
    \caption{Per-chunk playback DM.}
    \label{fig:gap}
  \end{subfigure}

  \vspace{-3mm}
  \caption{End-to-end QoE across network profiles.}
  \label{fig:preliminary-results}
  \vspace{-3mm}
\end{figure}

\para{Preliminary results and analysis.}
Figure~\ref{fig:preliminary-results} compares \sys with the baseline across increasingly constrained network profiles. As shown in Figure~\ref{fig:ttfr}, TTFR remains between 1.39 and 1.88~s across all profiles, compared with 1.77--5.41~s for the baseline. The improvement is modest under L1, at 21.3\%, but becomes substantially larger as network conditions deteriorate, reaching 74.4\% under L4. 
Notably, TTFR under L4 is lower than under L3, suggesting that the current scheduling policy responds differently across network regimes and leaves room for further exploration and tuning.

%More importantly, the metric remains relatively stable across profiles, whereas the baseline degrades sharply beyond L3. This suggests that cross-layer adaptation limits the accumulation of network delay and model-side backpressure along the end-to-end pipeline.
Figure~\ref{fig:gap} reports the average per-chunk lag beyond its playback deadline. \sys keeps this lag below 10~ms across all profiles, while the baseline increases from 46~ms under L1 to 94~ms under L4, indicating that chunks generated and delivered by \sys are more closely aligned with the intended playback timeline even under L4 constrained network conditions. Together, these results show that the design improves both initial responsiveness and sustained deadline adherence under network degradation.

\section{Discussion and Future Work}

\begin{figure}[t]
  \centering
  \includegraphics[width=0.48\textwidth]{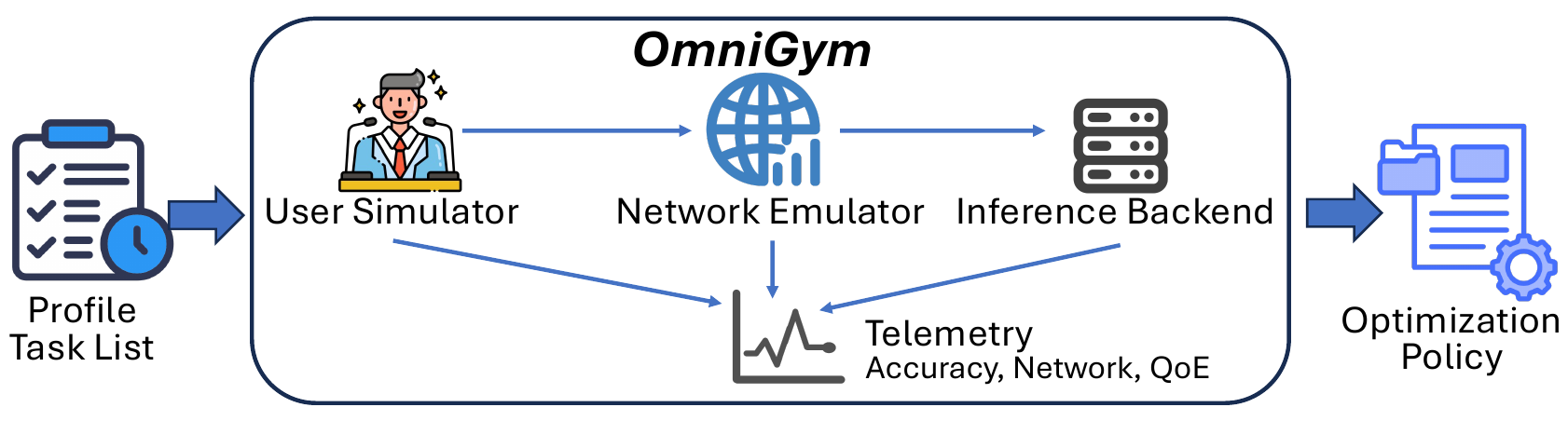}
  \vspace{-6mm}
  \caption{An envisioned OmniGym.}
  \vspace{-4mm}
  \label{fig:omnigym}
\end{figure}

%Since the networking community currently has limited understanding of full-duplex models and their impact on networks, a promising avenue for future research is to develop an \textit{OmniGym} that can systematically explore and assess optimization mechanisms for AI-native real-time communication.
\noindent\textbf{Exploring optimization policies with OmniGym.}
Since the networking community currently has limited understanding of full-duplex models and their impact on networks,
a promising direction is to build an \textit{OmniGym} for systematically exploring and evaluating optimization mechanisms for AI-native real-time communication. As shown in Figure~\ref{fig:omnigym}, \textit{OmniGym} could simulate diverse user interactions, emulate network and serving conditions, and perturb individual input modalities to evaluate their impact on model accuracy, grounding quality, turn-taking, and response latency. Such a platform would enable controlled and reproducible comparisons of alternative transport, scheduling, buffering, and modality adaptation policies before deploying them in real systems. It could also help identify effective configurations across different models, workloads, and operating conditions, thereby accelerating the design of model-aware end-to-end optimizations.

\para{Protocols for AI-native RTC.}
\sys currently schedules channels over existing protocols according to their transport characteristics, such as latency, reliability, ordering, and congestion-control behavior.
However, their heavyweight abstractions are difficult to adapt at the packet level for AI-specific behavior. In contrast, real-time AI communication is naturally chunk-based, with explicit timing, priority, modality, and validity semantics. This motivates a new lightweight UDP- or QUIC-based~\cite{quic} protocol that exposes such AI-native delivery semantics directly to the transport layer.

\para{Chunk-level serving engine observability.}
As full-duplex model serving shifts from request-level execution to continuous chunk processing, inference engines need fine-grained chunk-level observability and traceability. In particular, future systems should not only preserve the correspondence between input and output chunks, but also trace their execution across batching, modality encoding, LLM inference, and output generation, with a latency breakdown for each stage. 
Such visibility would enable accurate diagnosis of model backpressure and support chunk-aware scheduling across the serving, transport, and playback stages.

\para{Generalizing to other real-time AI media.}
Although \sys targets full-duplex omni-model interaction, its abstraction can be extended to other real-time AI media, including live-stream recommendation, real-time translation, meeting assistants, and interactive avatars. These workloads consume live multimodal streams under task-specific latency, fidelity, and alignment requirements. By expressing such requirements as model-centric specifications, the framework can combine them with network state and inference backpressure to make delivery and scheduling decisions across diverse applications.

% \section{CONCLUSION}

\section{Concluding Remarks}
Full-duplex AI-native interaction transforms sender-receiver real-time communication into a closed-loop system in which network transport, model inference, and user playback jointly determine interaction quality. This paper argues that model-facing chunks provide a common abstraction for coordinating these stages and presents \sys, an AI-native real-time communication framework that combines model-defined transport requirements, network- and session-aware inference scheduling, and adaptive receiver-side playback. Preliminary results demonstrate substantial QoE gains under constrained networks, highlighting the potential of jointly designing transport, serving, and playback architectures for real-time AI.

% \section*{Acknowledgements}
\newpage
\bibliographystyle{abbrv} 
\bibliography{citation}

\end{document}